# Open Science, Open Innovation? The Role of Open Access in Patenting Activity

Abdelghani Maddi[1] , Ahmad Yaman Abdin[2,3]  and Francesco De Pretis[*,4,5]

[1] Sorbonne Université, CNRS, Groupe d'Étude des Méthodes de l'Analyse Sociologique de la Sorbonne, GEMASS, F-75017 Paris, France

[2] Division of Pharmasophy, School of Pharmacy, Saarland University, 66123 Saarbrücken, Germany

[3] Division of Bioorganic Chemistry, School of Pharmacy, Saarland University, 66123 Saarbrücken, Germany

[4] Department of Environmental and Occupational Health, School of Public Health, Indiana University Bloomington, Bloomington, IN 47405, USA

[5] Department of Communication and Economics, University of Modena and Reggio Emilia, 42121 Reggio Emilia, Italy

*Corresponding author: Francesco De Pretis.

**Email:** francesco.depretis@unimore.it

**Author Contributions:** A.M. designed and performed research and wrote the paper. A.Y.A. contributed to writing, editing, and visualization. F.D. contributed to writing and project management.

**Competing Interest Statement:** Authors have no conflicts of interest to declare.

**Classification:** Social Sciences (major); Social Sciences (minor)

**Abstract**

Scientific knowledge is a key driver of technological innovation, shaping industrial development and policy decisions worldwide. Understanding how patents incorporate scientific research is essential for assessing the role of academic discoveries in technological progress. Non-Patent References (NPRs) provide a useful indicator of this relationship by revealing the extent to which patents draw upon scientific literature.

Here, we show that reliance on scientific research in patents varies significantly across regions. Oceania and Europe display stronger engagement with scientific knowledge, while the Americas exhibit lower reliance. Moreover, NPRs are more likely to be open access than the average scientific publication, a trend that intensifies when Sci-Hub availability is considered.

These results highlight the transformative impact of Open Science on global innovation dynamics. By facilitating broader access to research, Open Science strengthens the link between academia and industry, underscoring the need for policies that promote equitable and science-based innovation, particularly in developing regions.

**Introduction**

Innovation is widely recognized as a cornerstone of economic prosperity and societal progress [1–4]. Over the past decades, the global economy has increasingly relied on knowledge-intensive industries, where the ability to generate, absorb, and apply new knowledge determines competitiveness and long-term development [3,5–7]. At the heart of this transformation lies science—a tool for innovative problem-solving. Across sectors, from healthcare to renewable energy and beyond, the ability to innovate has multifaceted and multi-level impacts shaping both institutional success and economic trajectory of nations and regions [8].

Governments and policymakers increasingly emphasize fostering "innovation ecosystems", i.e. interconnected networks of research institutions, industries, funding bodies and regulations which drive knowledge transfer into technological advancement, so as to propel economic development and address societal challenges [9,10]. At the heart of this dynamic is the symbiotic relationship between science and innovation. Scientific research provides the foundational knowledge and breakthroughs which form the basis of technological advancements, driving productivity and economic gains [11,12]. The primordial role of science in economic and social progress has been articulated by the European Commission [13], which stated: "In the final years of the 20th century, we entered a knowledge-based society" and that "Economic and social development will depend essentially on knowledge in its different forms, on the production, acquisition, and use of knowledge". This shift has led to a growing recognition of the need to integrate science policies with economic strategies, aligning research investments with innovation-driven objectives. Consequently, measuring the extent to which industries rely on scientific research has become an essential step in evaluating the impact of science on economic activity.

In this context, Non-Patent References (NPRs) cited in patent documents have emerged as a key proxy for quantifying such a reliance, offering insights into the intersection of science and innovation [14–16]. Since the pioneering work of Carpenter and Narin (1983) [17] and Narin and Noma (1985) [18], NPRs have been widely analyzed to trace the knowledge sources which underlay technological advancements. Recent studies, such as those by Roach and Cohen (2013) [19] and Ahmadpoor and Jones (2017) [20], demonstrate the importance of NPRs in capturing knowledge flows from public research, i.e. publicly funded, to industrial innovation [10]. NPRs are particularly relevant in identifying the proximity of scientific disciplines to the technological frontier, with fields such as biotechnology and virology often cited in patents [20].

Furthermore, studies have distinguished between front-page and in-text NPRs. While front-page citations are frequently added for compliance with disclosure obligations, in-text NPRs are more closely tied to the inventive process [15,21,22]. This distinction emphasized the need to refine methodologies for measuring knowledge transfer, particularly as NPRs also serve to assess the impacts of public funding on innovation [14]. Recently, research has also demonstrated a substantial, though indirect, contribution of government-funded science to patentable technologies, particularly in fields such as clean energy and biomedical research [23,24]. Despite the growing body of literature, the extent to which openly accessible scientific research constitutes NPRs remains largely under explored [11]. This gap is particularly relevant given the growing emphasis on Open Science in policy, scientific practice, economic and societal impacts [25,26].

This article, therefore, seeks to fill this gap by exploring two interrelated research questions: 1) How does science contribute to innovation? And 2) To what extent is this contribution based on Open Science? The first question will explore the extent to which patents depend on scientific research via reporting the Reliance on Science Index (RSI) at different levels; regions (macro), technological domain (meso) and institution (micro). RSI is a metric which evaluates the extent to which patents depend on scientific knowledge by analyzing the proportion of NPRs cited within patent documents. The second question will employ the Normalized Open Access Indicator (NOAI) to investigate the openness of scientific research cited in such patents.

**Reliance on science: macro, meso and micro**

Figure 1A below illustrates the variation in Reliance on Science Index (RSI) by region. Regions such as Oceania and Europe exhibit median RSI values consistently above the neutral threshold of 1. Asia and Africa also show notable reliance, with medians exceeding 1, though with greater variability. The Americas experience a median RSI slightly below 1, indicating that patents from this region tend to rely less on scientific publications compared to others. Figure 1B, which examines the top 30 patent-producing countries, reveals the differences within regions. Countries such as South Korea, Taiwan, Singapore and Israel in Asia, and Denmark and Belgium in Europe, display particularly high RSI values. However, countries like South Africa, Mexico, Brazil, and Italy have RSI values below 1.

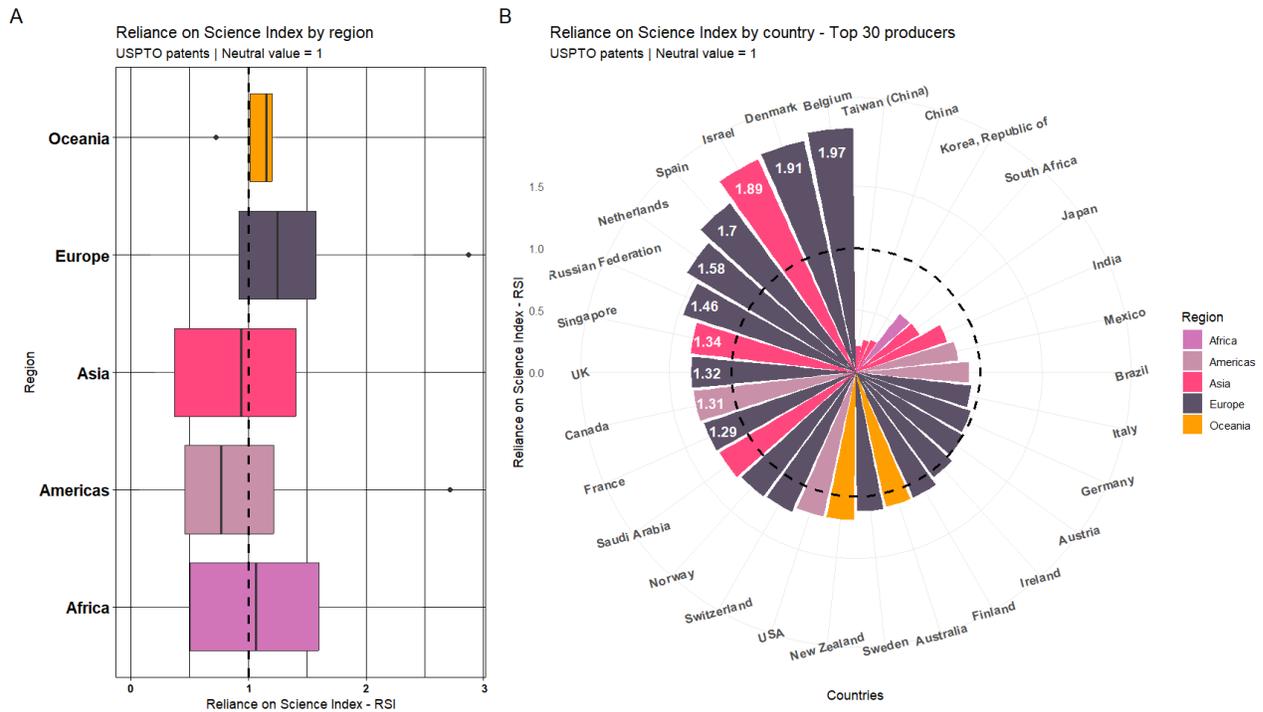

**Figure 1.** Reliance on science index by region (A), and by country – Top 30 producers (B). An RSI value greater than 1 indicates a higher-than-average reliance on scientific literature, while values below 1 suggest a weaker connection between science and innovation.

Table 1 identifies the top 10 International Patent Classification (IPC) with the highest RSI values. All classes in the list show values between 4 and 15, indicating that their reliance on scientific literature is 4 to 15 times higher than average.

**Table 1.** IPC classes with highest Reliance on Science Index

| IPC Code | IPC Label | Example of R&D / inventions | # Patents citing NPRs | Share in patents citing NPRs (%) | Reliance on Science Index - RSI |
|---|---|---|---|---|---|
| C07K | Peptides | Peptide-based cancer vaccines; Insulin analogs for diabetes management | 17,590 | 3.54% | 12.34 |
| C07H | Sugars; Derivatives thereof; Nucleosides; Nucleotides; Nucleic acids | CRISPR technology for genetic editing; Synthetic DNA sequences for vaccines | 10,725 | 2.16% | 11.08 |
| C12N | Microorganisms or enzymes; Mutation or genetic engineering; Tissue culture | Genetically modified bacteria for insulin production; Microbial strains for probiotic products | 30,575 | 6.15% | 9.95 |
| C12P | Fermentation processes or processes using enzymes for synthesizing chemical compounds | Production of ethanol from corn using yeast; Enzyme-assisted lactose-free milk production | 10,263 | 2.06% | 9.74 |
| C12Q | Measuring or testing processes involving enzymes or microorganisms | COVID-19 diagnostic test kits; Microbial contamination sensors for food safety | 14,915 | 3.00% | 9.29 |
| A61K | Preparations for medical, dental, or hygienic purposes | Antibiotics for bacterial infections; Anti-inflammatory ointments | 67,969 | 13.67% | 6.45 |
| A01N | Preservation of human or animal bodies or plants; Biocides; Attractants or repellents for animals; Plant growth regulators | Insecticides for crop protection; Antifungal agents for plants | 7,316 | 1.47% | 4.61 |
| C07D | Heterocyclic compounds | Development of antiviral drugs containing nitrogen heterocycles; Anti-inflammatory medications | 19,763 | 3.98% | 4.54 |
| C07F | Acyclic, carbocyclic, or heterocyclic compounds containing elements other than carbon, hydrogen, halogen, oxygen, nitrogen, sulfur, selenium, or tellurium | Development of organometallic catalysts for polymerization; Silicon-based compounds for semiconductors | 3,240 | 0.65% | 4.19 |
| A61P | Specific therapeutic activity of chemical compounds or medicinal preparations | Chemotherapy drugs for cancer treatment; Antiviral medications for HIV | 3,152 | 0.63% | 4.17 |

Biotechnology-related IPC classes, such as C07K (Peptides), C07H (Nucleic Acids and Sugars), and C12N (Microorganisms and Genetic Engineering), exhibit some of the highest Reliance Index values (12.34, 11.08, and 9.95, respectively). These fields rely extensively on scientific research to develop technologies such as peptide-based therapies, CRISPR gene-editing tools, and genetically engineered bacteria. The elevated RSI values reflect a significant role for academic studies and experimental research in advancing these domains, where the transition from discovery to application is often direct. For instance, innovations in C12N, such as microbial strains used in pharmaceuticals, naturally depend on basic scientific research. Similarly, C07H encompasses synthetic DNA sequences essential for vaccine development, highlighting how such advancements build upon a well-established body of scientific work [27].

In contrast, more applied or practical sectors such as B42F (filing appliances) or E02F (dredging) exhibit much lower RSI values, Table 2. Such industries, which focus on optimizing and innovating around existing technologies or improving manufacturing processes, have less direct reliance on fundamental scientific research. Their innovation is more likely to stem from engineering improvements, cost efficiency, and incremental technological advancements. For instance, in IPC classes like B25C (hand-held tools) or E06C (ladders), innovations are often more focused on product usability, ergonomic designs, or incremental material improvements, with less emphasis on groundbreaking scientific research.

**Table 2:** IPC classes with the lowest Reliance on Science Index

| IPC Code | IPC Label | Example of R&D / inventions | # Patents citing NPRs (fractional count) | # All US Patents (fractional count) | Reliance on Science Index - RSI |
|---|---|---|---|---|---|
| E02F | Dredging; Soil-shifting | Development of autonomous dredging systems for coastal engineering projects. | 0.3 | 2,163.51 | 0.002 |
| B42F | Sheets temporarily attached together; Filing appliances | Innovative binders and filing systems for office and educational use. | 0.3 | 2,056.32 | 0.002 |
| E06C | Ladders | Foldable, lightweight ladders with integrated safety mechanisms. | 0.3 | 1,969.82 | 0.002 |
| B25C | Hand-held nailing or stapling tools | Ergonomic, battery-powered nail guns for construction. | 0.5 | 3,317.64 | 0.003 |
| D05B | Sewing | Advanced sewing machines with AI-assisted stitching for complex patterns. | 0.8 | 5,297.84 | 0.003 |
| H10B | Electronic memory devices | Development of non-volatile memory for high-speed data storage. | 0.3 | 1,922.64 | 0.003 |
| D06F | Laundering, drying, ironing, pressing or folding textile articles | Smart laundry systems integrating fabric-specific drying and folding technologies. | 0.5 | 2,337.59 | 0.003 |
| E05F | Devices for moving wings into open or closed position; Checks for wings; Wing fittings not otherwise provided for, concerned with the functioning of the wing | Automatic sliding door systems with energy-efficient sensors. | 0.3 | 1,259.87 | 0.004 |
| F01L | Cyclically operating valves for machines or engines | High-durability valves for industrial machinery with improved flow regulation. | 0.3 | 1,163.86 | 0.004 |
| E03F | Sewers; Cesspools | Eco-friendly sewage treatment systems with integrated filtration. | 0.2 | 714.63 | 0.005 |
| E03C | Domestic plumbing installations | Smart plumbing systems with water-saving features and leak detection. | 0.6 | 1,948.6 | 0.005 |

Table 3A illustrates the analysis of the top 20 organizations listed based on RSI. The values exceeding 11 reflect the role of basic research in driving advancements across various sectors and institutions. Biopharmaceutical companies dominate this list, with institutions such as Ionis Pharmaceuticals, Incyte Pharmaceuticals, and Biogen Idec achieving the highest indices, ranging from 13 to 15. These organizations rely heavily on cutting-edge scientific research in fields such as genetic engineering, peptide-based therapies, and molecular biology to develop novel drugs and therapies.

**Table 3.** Organizations with highest (A) and lowest (B) Reliance on Science Index

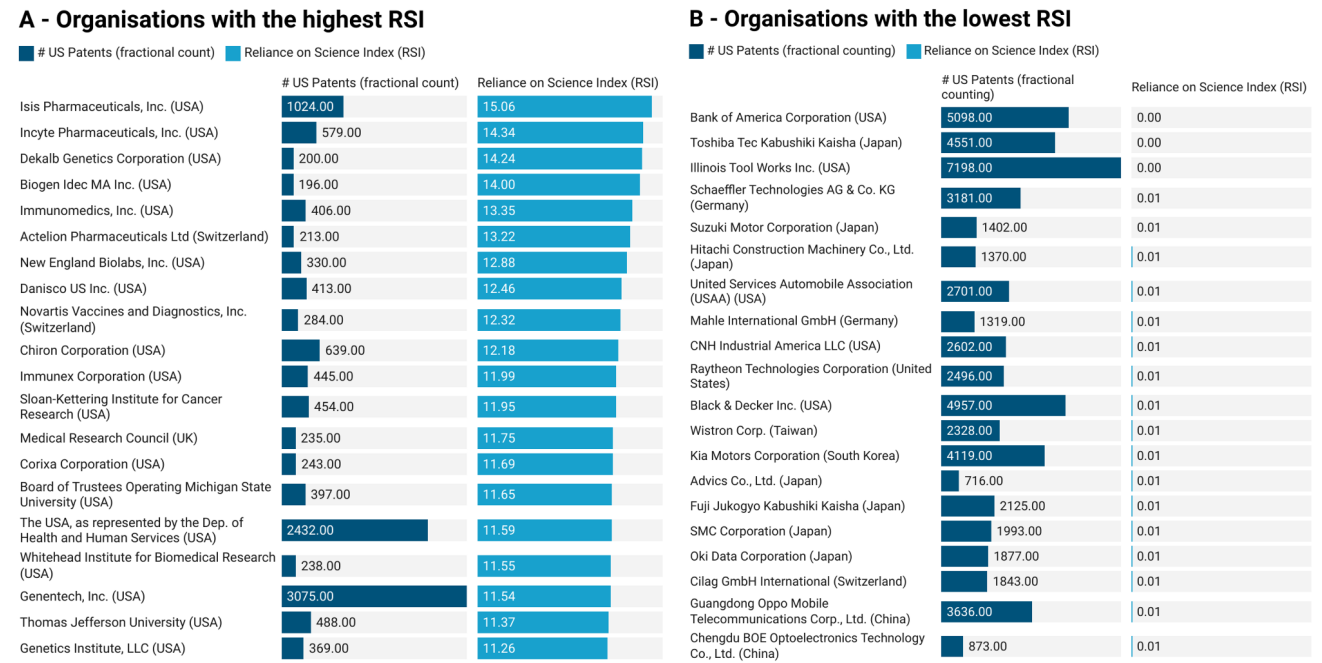

on academic discoveries to develop emerging technologies. These companies operate in niche markets, producing tools and solutions for broader applications in biotechnology and agriculture.

**Openness and disciplinary characterization of NPRs**

The share of OA publications in NPRs is shown in Figure 2A. Surprisingly, non-OA citations account for 71% of NPRs. While this might initially seem counterintuitive, the fact that 29% of NPRs are OA is still a substantial proportion. This aligns with the reality that, despite global efforts to promote Open Access, OA publications still constitute less than 50% of the overall scientific literature across various disciplines [28]. Additionally, our analysis is based on data from OpenAlex, which is a comprehensive database, though other sources may yield slightly different results. Interestingly, when black OA sources such as SciHub are included, the share of OA citations rises dramatically to 85%, highlighting the significant role of accessibility in shaping NPR patterns. The broader impact of OA on NPRs is better understood through the NOAI as it assesses the relative prevalence of OA publications within NPRs compared to their prevalence in the broader scientific corpus, in our case OpenAlex database.

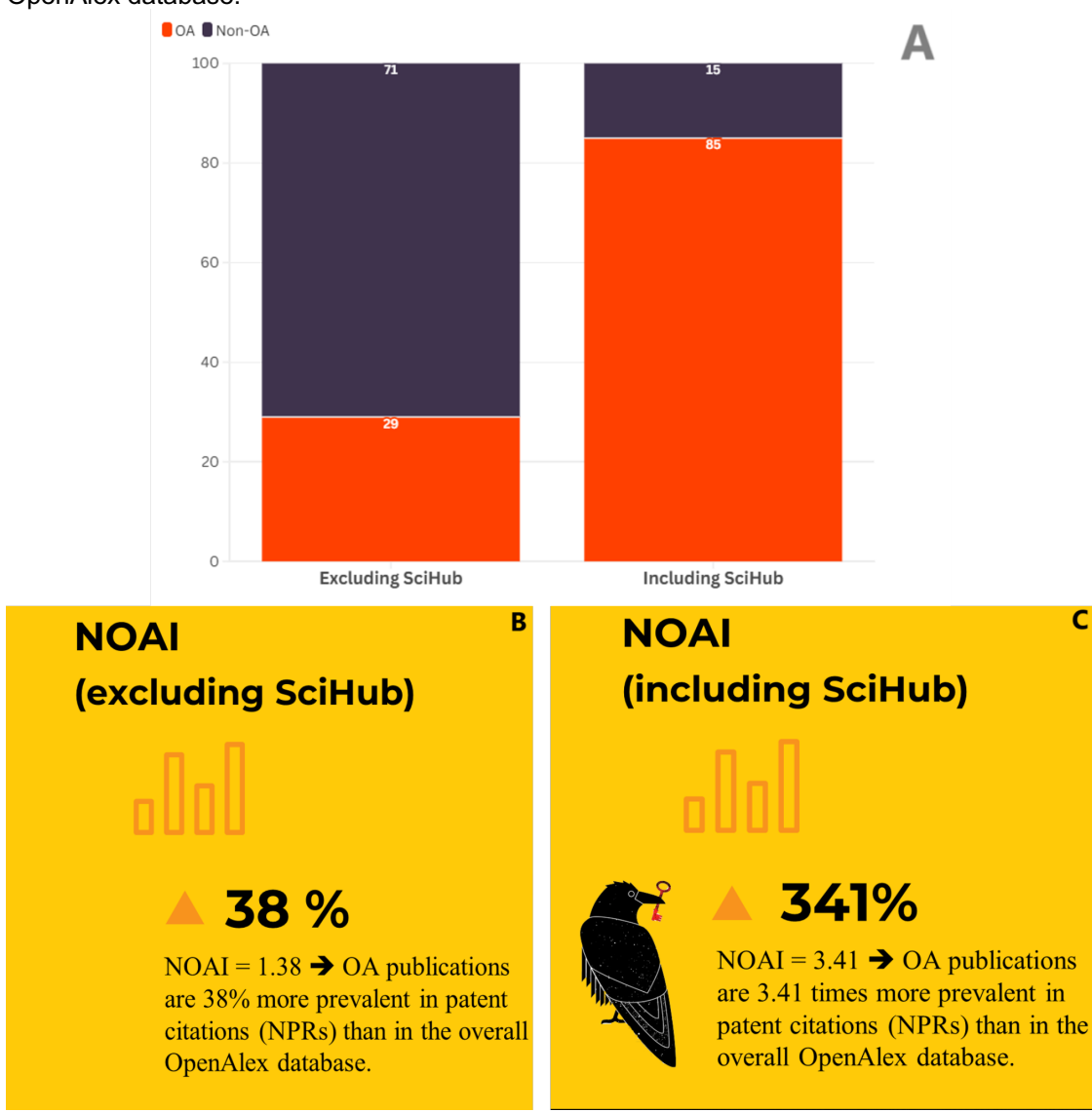

**Figure 2.** Share of OA publications in NPRs (A), NOAI excluding SciHub (B), and NOAI including SciHub (C).

The analysis of the Normalized OA Index (NOAI) in Figure 2B indeed provides additional insights. Firstly, The overall NOAI value of 1.38 indicates that OA publications are 38% more prevalent in NPRs when compared to the broader OpenAlex database. This suggests that OA literature is actually particularly favored in patent citations, likely due to its accessibility and relevance to innovation. Secondly and perhaps more interestingly, when factoring in sources available through black OA, as shown in Figure 2C, the accessibility of NPRs increases dramatically to more than three times.

The analysis of the distribution of Non-Patent References (NPRs) by discipline, is shown in Figure 3A. On one side, disciplines such as Biology (27.14%), Chemistry (23.49%), and Medicine (16.34%) dominate in terms of their share of NPRs. This aligns with their inherent strong transferability and industrial applicability, particularly in areas such as pharmaceuticals, biotechnology, and chemical engineering. On the other side, disciplines such as art, history, geography, and sociology, each accounting for less than 1% of NPRs, are less represented. This distribution reflects the practical and applicable orientation of NPRs, which tend to be cited in patents relevant to technological or industrial innovations. Furthermore, the discipline based NOAI values are shown in Figure 3B. The differences may reflect either the degree of availability of OA publications in these fields or differences in their citation practices. Computer science and business, for instance, show higher-than-average NOAI values, despite their relatively modest shares of OA publications in the NPR dataset. This may indicate a selective but impactful use of OA resources, where open literature is leveraged for specific innovation-driven purposes. Interestingly, when availability via black OA is included, the NOAI values exceed 1 across all disciplines, Figure 3C. This suggests that even in fields with lower formal OA integration, the practical availability of research literature is significantly enhanced through black OA.

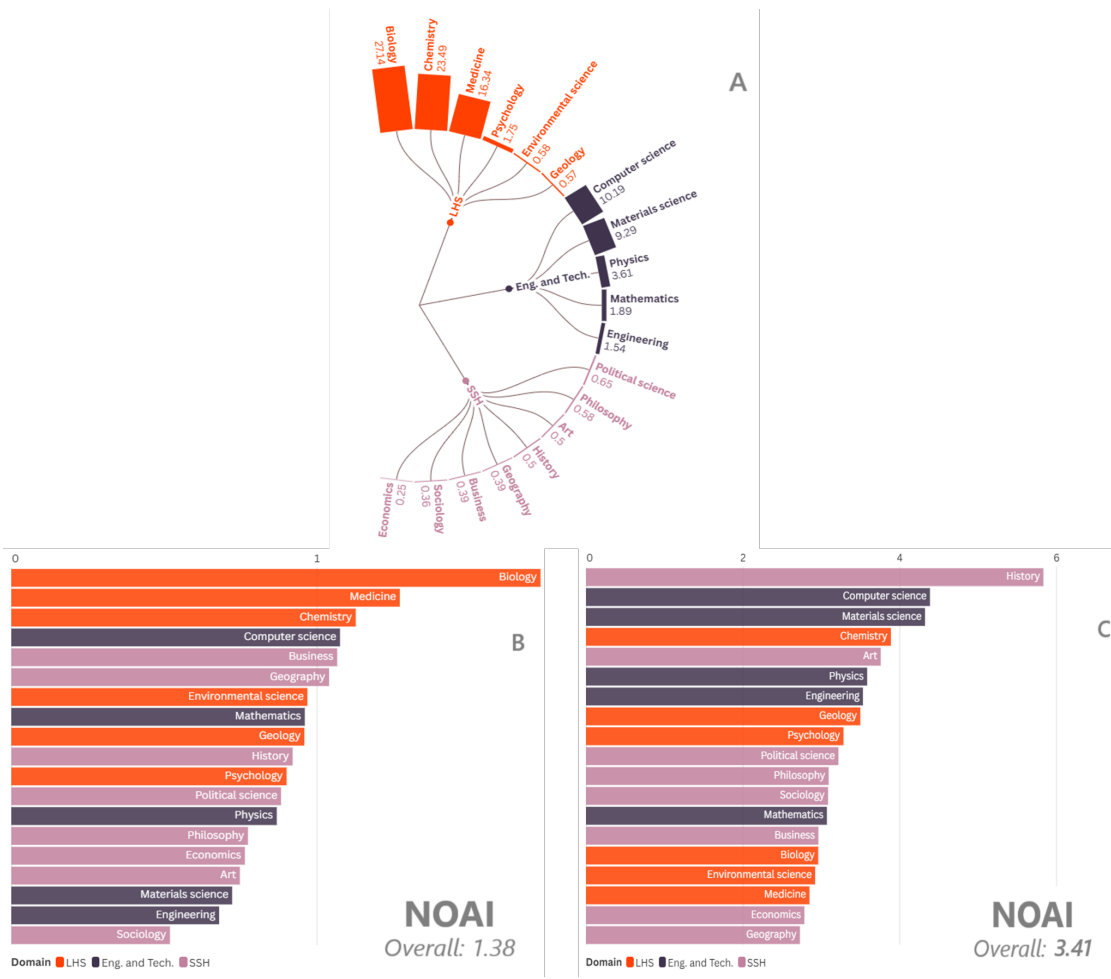

**Figure 3.** Distribution of NPRs by discipline (A), and the NOAI with (B) and without SciHub (C).

**Discussion**

The findings of this study portray a comprehensive picture of the relationship between basic research and innovation. The reliance of patents on scientific knowledge is examined across macro (region and country), meso (technological domain), and micro (organization or institution) levels.

Subsequently, the representation of Open Science is measured. Insights into these dynamics form an invaluable foundation for reflecting on the policy and strategic implications for both research practices and innovation.

At the macro level, the analysis by regions and countries reveals notable differences in the integration of science into innovation activities and aligns with previous research [29]. Advanced economies, particularly the United States and Western Europe, demonstrate a strong reliance on science, evidenced by the substantial presence of Non-Patent References (NPRs) in their patents. This can be attributed to well-funded research systems, a culture of collaboration between public and private sectors, and policies promoting access to research outputs. Conversely, some emerging or developing regions show lower utilization of NPRs, indicating challenges related to access to research, R&D capacity, or scientific infrastructure. These disparities underline the importance of global initiatives aimed at bridging the gap in access to and use of scientific knowledge, particularly to foster inclusive and equitable innovation.

At the meso level, the examination of industrial sectors and IPC (International Patent Classification) classes reveals distinct patterns of scientific literature usage across technological domains. Sectors such as biotechnology, pharmaceuticals, and life sciences overwhelmingly dominate, confirming their heavy reliance on fundamental and applied research. This aligns with the innovation demands in these fields, where scientific discoveries often directly underpin the development of products and technologies. In contrast, sectors such as mechanical engineering or electronics exhibit a more moderate reliance on science, reflecting a focus on engineering or practical knowledge less directly derived from academic research. These differences highlight the need for tailored strategies to support innovation, with a particular emphasis on disciplines and sectors where collaboration between science and technology is critical.

At the micro level, the analysis of organizations reveals diversity in how individual entities—companies, universities, and public institutions—integrate science into their patents. Leading pharmaceutical and biotech companies, such as Genentech and Isis Pharmaceuticals, stand out with exceptionally high reliance on science indices, probably pointing towards their position at the forefront of scientific innovation. Simultaneously, public and academic institutions, such as the Sloan-Kettering Institute or the National Institutes of Health (NIH), play a crucial role as sources of scientific knowledge for industry. These results emphasize the importance of public-private partnerships and policies that support both fundamental research and its transfer to industrial applications.

Open Access publications are indeed overrepresented in patents. This advantage translates into the practical utility of scientific knowledge supporting both economic prosperity and societal progress. The Normalized OA index (NOAI) demonstrates that scientific knowledge available through open access is, on average, 38% more likely to be cited by patents. If black open access, i.e. publications available via SciHub, is considered, the overall NOAI value dramatically jumps to 341%. While scientific field-dependent discrepancies exist, the trends observed follow those of the RSI at the meso level. In other words, patents from technological domains which rely more on science are likely to cite open access literature and this relationship is increased when barriers to accessibility are further removed. Taken together, these findings suggest that the integration of Open Science practices is indeed a significant driver of innovation.

Here, it is also worth noting that a rather significant proportion of basic research integrated in patented innovations is produced within publicly funded institutions. This may raise critical, and also ethical, considerations regarding the equitable distribution of the economic value generated through innovation. While industries benefit from the accessibility of publicly financed research, the mechanisms for reinvesting these gains into the broader scientific ecosystem remain uneven. Addressing this issue through policies which focus on the reciprocity between publicly funded research and private sector innovation could enhance the sustainability of the knowledge economy.

There is a compelling necessity to develop, adopt and implement holistic policies which account for regional, sectoral and organizational specificities. Enhancing accessibility to scientific knowledge and fostering broader integration of underrepresented disciplines and regions, may indeed maximize the impact of science on addressing urgent societal and technological challenges. Future studies ought to focus on further explaining the effect of OA policies on Innovation ecosystems and perhaps looking in more depth into the economic, social as well as environmental aspects of knowledge transfer.

**Materials and Methods**

We employed two comprehensive datasets: patent data from the United States Patent and Trademark Office (USPTO), as provided by PatentsView, and scientific citation data from the Marx et al. (2023) database [30]. The patent data includes over 22.5 million inventors associated with 8.9 million patents, along with detailed metadata on 96,039 disambiguated locations and 8.2 million organizations. The citation data allows us to focus specifically on NPRs cited by inventors in the text or body of patents, encompassing 3.2 million NPRs cited by 497,098 patents and representing nearly one million distinct scientific publications.

To address our research questions, we employed two complementary indicators: the Reliance on Science Index (RSI) and the Normalized Open Access Index (NOAI). These indicators were specifically designed to analyze the relationship between patents and scientific literature, providing insights at multiple levels—macro (regions and countries), meso (sectors and classes), and micro (organizations).

The RSI quantifies the relative reliance of a given category (e.g., region, country, sector, or organization) on scientific literature, using a fractional count approach where the total contribution of a single patent to the indicator sums to 1. It is calculated as:

$$\text{RSI}_i = \frac{\#\text{Patents citing NPRs}_i}{\#\text{Patents in all NPRs corpus}} \Big/ \frac{\#\text{Patents}_i \text{ (at USPTO)}}{\#\text{Patents in all USPTO}}$$

Here, *i* represents the category under consideration. The numerator captures the share of patents citing NPRs in category i relative to the total patents citing NPRs in the entire corpus. The denominator represents the share of all patents from category iii relative to the total patents in the USPTO dataset. An RSI greater than 1 indicates a higher-than-average reliance on scientific literature, while an RSI below 1 signals a lower reliance.

The NOAI, meanwhile, assesses the relative prevalence of Open Access (OA) publications within NPRs compared to their prevalence in the broader scientific corpus, namely OpenAlex database in this study [31]. This measure reflects the extent to which patent citations incorporate accessible scientific knowledge. For the detailed formula, refer to Maddi (2020) [32]. A NOAI greater than 1 suggests that OA publications are overrepresented in NPRs, underlining their importance in the innovation process.

**Acknowledgments**

This work was presented at *the Nordic Workshop on Bibliometrics & Research Policy 2024* conference in Reykjavík – Iceland, and we would like to thank our colleagues for their valuable suggestions, which significantly enhanced the quality of this analysis. The presentation is available at the following link: https://zenodo.org/records/14205167